\documentclass{article}
\usepackage{spconf,amsmath,graphicx, amsfonts, mathtools}

\title{TOWARDS ADVERSARIAL LEARNING OF SPEAKER-INVARIANT REPRESENTATION FOR SPEECH EMOTION RECOGNITION}
\name{Ming Tu, Yun Tang, Jing Huang, Xiaodong He, Bowen Zhou}
\address{JD AI Research \\
          675 East Middlefield Road \\
          Mountain View, CA 94043, USA}
\begin{document}
%
\maketitle
\begin{abstract}
Speech emotion recognition (SER) has attracted great attention in recent years due to the high demand for emotionally intelligent speech interfaces. 
Deriving speaker-invariant representations for speech emotion recognition is crucial. In this paper, we propose to apply adversarial training to SER to learn speaker-invariant representations. Our model consists of three parts: a representation learning sub-network with time-delay neural network (TDNN) and LSTM with statistical pooling, an emotion classification network and a speaker classification network. Both the emotion and speaker classification network take the output of the representation learning network as input. Two training strategies are employed: one based on domain adversarial training (DAT) and the other one based on cross-gradient training (CGT). Besides the conventional data set, we also evaluate our proposed models on a much larger publicly available emotion data set with 250 speakers. Evaluation results show that on IEMOCAP, DAT and CGT provides 5.6\% and 7.4\% improvement respectively, over a baseline system without speaker-invariant representation learning on 5-fold cross validation. On the larger emotion data set, while CGT fails to yield better results than baseline, DAT can still provide 9.8\% relative improvement on a standalone test set.
\end{abstract}
\section{Introduction}
\label{sec:intro}

With intelligent speech assistants such as Alexa, Google Home, Siri and Cortana being used in our daily lives more than ever, we still notice the performance gap of these machine dialogues from human interactions, because these systems lack of capability of recognizing our emotions and react to them like a human partner would. Therefore, the demand is rising for speech emotion recognition (SER) to empower dialogue systems to respond in emotionally intelligent ways, especially for customer service chatbots. However, SER is challenging due to mismatches between training and testing data in terms of speaker variations, recording environment/channels etc. 
It is impractical to train one SER system which could cover every application scenario.

Various methods have been proposed to tackle the negative effect of data variations and domain mismatch in SER literature. Some studies resort to extracting richer features representing emotion variation or applying more powerful deep neural networks (DNN) with careful architecture designs, expecting better generalization ability to unseen scenarios  \cite{chen20183,luo2018investigation} . 
Another research direction explicitly deals with the domain mismatch problem by compensating the data variations with either robust feature learning or model/training strategy design \cite{schuller2010cross,abdelwahab2017ensemble}. Ideas of domain adaptation in general machine learning literature are borrowed to help similar tasks in the SER field.

Generative Adversarial Networks (GAN) has achieved much success in speech applications \cite{pascual2017segan, kaneko2017generative,hsu2017voice,jiao2018simulating}. Domain adaptation techniques developed from GANs have also been applied to common domain mismatch problems, such as: automatic speech recognition (ASR) \cite{shinohara2016adversarial, sun2017unsupervised, meng2018speaker}, cross-corpus speaker recognition \cite{wang2018unsupervised} and SER \cite{abdelwahab2018domain}. All these studies are based on the domain adversarial training (DAT) proposed in \cite{ganin2016domain}.
DAT was applied to unsupervised domain adaptation by training the main task and domain classifier at the same time. A gradient reversal layer (GRL) was inserted to the domain classifier in order to confuse the domain classifier while accomplish the main task well. The representation learning in this way is more robust to domain shifts and variations, which have been proved in the aforementioned studies. 

In this work, we propose to apply adversarial training to SER. 
In order to deal with speaker variations, this paper aims to learn speaker-invariant representations for SER, and expects the representation learning network to generalize well to unseen testing speakers. Our model consists of three parts: a representation learning network with time-delay neural network (TDNN) and LSTM with statistical pooling layer (a variation of the popular x-vector \cite{snyder2017deep} for speaker recognition in Kaldi \cite{povey2011kaldi}), an emotion classification network and a speaker classification network. Both the emotion and speaker classification networks take the output of representation learning network as input. Two adversarial training strategies are employed to achieve speaker-invariant representations: one based on the original DAT \cite{ganin2016domain}, and the other one based on cross-gradient training (CGT)\cite{shankar2018generalizing}.  In contrast to using a GRL in DAT, CGT develops a domain-guided perturbation that serves as data-augmentation during training, and proves this training strategies can generalize well to unseen domains. 
We evaluate the proposed systems on two data sets: one is the commonly used Interactive Emotional Dyadic Motion Capture (IEMOCAP) data set, and the other one is a public available Mandarin speech emotion data set collected by SpeechOcean. It has 250 speakers (much more than IEMOCAP) in total. Evaluation results show that on IEMOCAP, DAT and CGT respectively provides 5.6\% and 7.4\% improvement over a baseline SER system without speaker-invariant representation learning on 5-fold cross validation. On the larger Mandarin speech emotion data set, while CGT fails to yield better results than baseline, DAT can still provide 9.8\% relative improvement on a standalone test set.

\textbf{Relation to prior work:}
Our work in this paper is related to previous studies on using domain adaptation based on DAT for speech applications \cite{shinohara2016adversarial, sun2017unsupervised, wang2018unsupervised, abdelwahab2018domain}. These studies mainly focus on cross-domain tasks and solve it as a domain adaptation problem. These work requires unlabeled target domain data to do adversarial training. Our primary task, however, is to derive speaker-invariant representation for SER, and expect it can generalize well to unseen speakers (more like a domain generalization problem). Furthermore, we do not assume target speaker data (either labeled or unlabeled) is available during training.
Different from \cite{meng2018speaker} which uses adversarial training to derive speaker-invariant features for ASR, we also employ another training strategy CGT besides DAT. Instead of using GRL to achieve adversrial learning, CGT uses the gradients of adversarial tasks to guide perturbations on input to achieve domain generalization\cite{shankar2018generalizing}. To the best of our knowledge, this paper is the first work to apply CGT on SER. Our work is also different from previous work on domain adaptation/generalization for SER or cross-corpus SER \cite{schuller2010cross,abdelwahab2017ensemble}, in that we employ the adversarial training strategies to achieve generalization to unseen speakers.

\section{SYSTEMS DESCRIPTION}
\label{sec:sys}

Assume data set $\{ (\mathbf{X}_i, \mathbf{y}_i, \mathbf{s}_i) \}^{N}_{i=1}$, where $\mathbf{X}_i \in \mathbb{R}^{d \times l_i}$ is the sequence of Mel Frequency Cepstral Coefficients (MFCC) of utterance $i$ ($d$ is the MFCC dimension and $l_i$ is the sequence length). $\mathbf{y}_i$ and $\mathbf{s}_i$ are the one-hot encoded emotion label and speaker label of utterance $i$. Our goal is to derive a fixed-length embedding $\mathbf{e}_i $ for utterance $i$, which is supposed to be able to well discriminate speech emotions while be irrelevant with speaker identities, i.e., to train an embedding network for SER that is robust to unseen speakers.

\subsection{Model architecture}
\begin{figure}[t]
   \centering
   \includegraphics[width=0.7\linewidth]{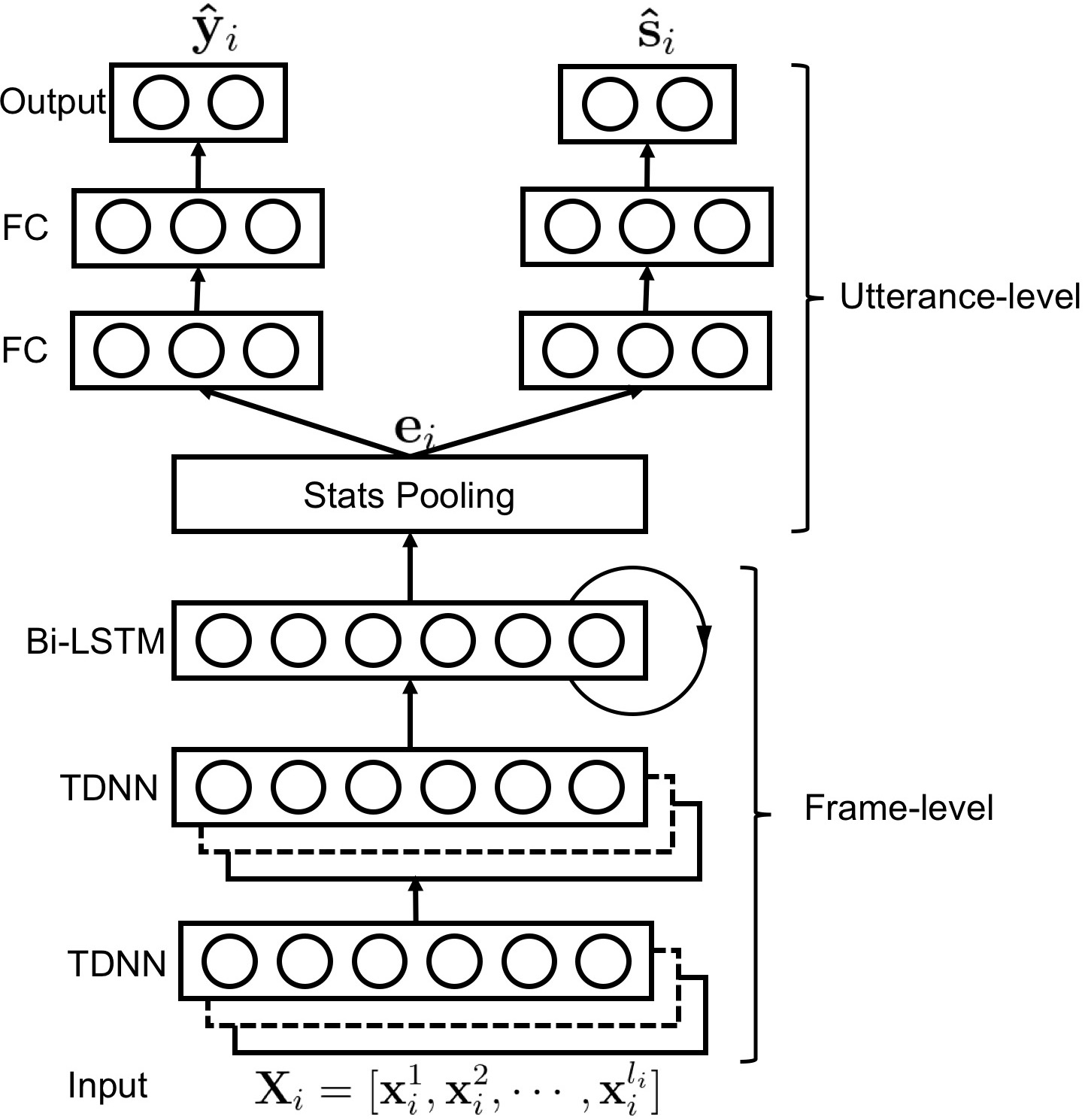} 
   \caption{Model architecture}
   \label{fig:fig1}
\end{figure}

To achieve speaker-invariant embeddings, we propose a network with multi-task learning setting to utilize both emotion and speaker labels of training set. Figure \ref{fig:fig1} illustrate the network architecture in this study. The model consists of three sub-networks. The embedding sub-network takes sequence of MFCC features $\mathbf{X}_i = [\mathbf{x}^1_i, \mathbf{x}^2_i, \cdots, \mathbf{x}^{l_i}_i]$ as input. Two layers of TDNN and 1 layer of bidirectional recurrent neural network (RNN) with Long Short Term Memory (LSTM) nodes are in charge of sequential feature extraction. 
Then, mean and standard deviation are calculated along the sequence and concatenated to derive a fixed-dimension utterance-level embedding $\mathbf{e}_i$, which is input to two following sub-networks to accomplish corresponding classification tasks. The left sub-network outputs predicted emotion label $\mathbf{\hat{y_i}}$ and the right sub-network outputs predicted speaker label $\mathbf{\hat{s_i}}$. Both of them are fully connected networks with two hidden layers. Without speaker recognition (SR) sub-network, it becomes a single-task learning SER system; With normal DNN training strategies for both of SER and SR, it becomes a multi-task learning system \cite{zhang2017cross}. Next subsections will introduce two adversarial training strategies for the same model architecture.

\subsection{Domain adversarial training (DAT)}

The key idea of DAT is to treat to $\mathbf{s}_i$ as domain label. Then the right sub-network in figure \ref{fig:fig1} becomes a domain classifier. In order to learn domain invariant representations between source and target domain, DAT inserts a gradient reversal layer (GRL) to reverse the gradient flow from right sub-network to the embedding sub-network by multiplying the gradients there by a negative value $\lambda$; while in forward pass, there is no such effect and the GRL just works like identity mapping. In this way, $\mathbf{e}_i$ tends to confuse the SR sub-network and still can predict the emotion label well. The loss of DAT can be defined as:

\begin{math}
\begin{aligned} 
   L(\mathbf{\Theta}) = \sum_{i=1}^{N}( L(\mathbf{y_i}, \mathbf{\hat{y_i}}; \mathbf{X}_i, \mathbf{\Theta}) - \lambda L(\mathbf{s_i}, \mathbf{\hat{s_i}}; \mathbf{X}_i, \mathbf{\Theta})) \quad (1)
\end{aligned}
\end{math}

where $\mathbf{\Theta}$ is the model parameters, and $L(\mathbf{y_i}, \mathbf{\hat{y_i}}; \mathbf{X}_i, \mathbf{\Theta})$ is the loss of emotion classifier (same for the speaker classifier). Original DAT requires unlabelled data from or similar to target domain to achieve domain adaptation. Previous study either use recording condition or corpus identity as domain labels to deal with environment or corpus variation. This paper applies DAT to learn speaker-invariant representations by jointly adversarial training on training data with both emotion and speaker labels.

\subsection{Cross-gradient training (CGT)}

CGT is proposed in \cite{shankar2018generalizing}  to solve domain generalization problem. It frees the requirement for target domain data during training, and develops a scheme that can generalize to unseen domains during test. Instead of aiming at reducing domain specific information in $\mathbf{e}_i$, CGT introduces domain-guided perturbations of the input based on gradients of adversarial tasks. Then, the emotion recognition sub-network can be trained on both the original input and the input produced by the data augmentation during batch training. The domain-guided perturbation on inputs makes the model cover more domain variations and robust to potential domain shifts during testing. The basic training procedure of CGT is summarized as follows:

\begin{math}
\begin{aligned} 
   & \mathbf{X}^s_i := \mathbf{X}_i + \epsilon \nabla_{\mathbf{X_i}}  L(\mathbf{s_i}, \mathbf{\hat{s_i}}; \mathbf{X}_i, \mathbf{\Theta}) \qquad \qquad \qquad \qquad \, (2)\\
   & \mathbf{X}^y_i := \mathbf{X}_i + \epsilon \nabla_{\mathbf{X_i}}  L(\mathbf{y_i}, \mathbf{\hat{y_i}}; \mathbf{X}_i, \mathbf{\Theta}) \qquad \qquad \qquad \quad \;\;\; (3)
\end{aligned} \\
\begin{aligned}
   \mathbf{\Theta} \leftarrow & \mathbf{\Theta}-\eta\nabla_{\mathbf{\Theta}} ( (1-\alpha) ( L(\mathbf{y_i}, \mathbf{\hat{y_i}}; \mathbf{X}_i, \mathbf{\Theta}) + L(\mathbf{s_i}, \mathbf{\hat{s_i}}; \mathbf{X}_i, \mathbf{\Theta})) \\
   & + \alpha ( L(\mathbf{y_i}, \mathbf{\hat{y_i}}; \mathbf{X}^s_i, \mathbf{\Theta}) + L(\mathbf{s_i}, \mathbf{\hat{s_i}}; \mathbf{X}^y_i, \mathbf{\Theta})) ) \quad \qquad \;\;\; (4)
\end{aligned}
\end{math}

where $\mathbf{X}^s_i$ and $\mathbf{X}^y_i$ are the perturbed inputs. Eq. 4 gives the parameter update formula. It has been shown in \cite{shankar2018generalizing} that CGT is more stable and easier to train than DAT.

\section{EXPERIMENTAL SETUP}
\label{sec:print}

\subsection{Data preparation}
\textbf{IEMOCAP:} IEMOCAP \cite{busso2008iemocap} was collected in 5 sessions, each of which has one female and one male speaker in both scripted and improvised scenarios. Categorical emotion annotations include 9 classes. In this paper, we only use the improvised recordings. We put \textit{happy} and \textit{excitement} into one class to achieve more balanced label distribution. There are totally 4 emotion classes \{\textit{happy}, \textit{sad}, \textit{angry}, \textit{neutral}\} and 2943 utterances. In order to get more reliable evaluation, 5-fold cross validation is employed with 4 sessions for training, one speaker in the remaining session for validation and one speaker for testing.

\textbf{Mandarin speech emotion data set:} This data set was collected by SpeechOcean with 250 recruited speakers. Each speaker was asked to read 240-260 sentences in Mandarin with four emotions \{\textit{happy}, \textit{sad}, \textit{angry}, \textit{surprise}\} together with \textit{laughing} and \textit{cry}. The four emotion classes were balanced to ensure there were almost the same number of utterances for each emotion class.
Mobile phones belong to different brands and equipped with different operating systems were used as recorder. This data set is suitable for speech emotion recognition research for the reason that it has much more speakers and utterances than all available speech emotion data sets in literature; it also has very balanced emotion class distributions. For this study, 53803 utterances (durations: 2.80$\pm$0.68 seconds) with four emotions from all 250 speakers are employed. We randomly picked 200 speakers (43202 utterances) for training, 25 speakers (5407 utterances) for validation and 25 speakers (5194 utterances) for testing.

\textbf{Feature extraction:} 13-dimensional MFCC together with first and second order derivatives are extracted with Kaldi. Simple energy-based voice activity detection is used to remove silences to relieve its impact on statistics pooling. Finally, each utterance is with a sequence of 39-dimensional MFCC features, an emotion label and a speaker label. For both data sets, speakers in validation and testing set are never seen in training.

\begin{table}[t]
\centering
\caption{Model configurations for two data sets.}
\resizebox{0.95\columnwidth}{!}{%
\begin{tabular}{|c|c|c|}
\hline
 & IEMOCAP & Mandarin SE dataset \\ \hline
\begin{tabular}[c]{@{}c@{}}Embedding\\ sub-network\end{tabular} & \begin{tabular}[c]{@{}c@{}}TDNN 128-5-2\\ TDNN 64-3-4\\ Bi-LSTM 64\\ FC 256\end{tabular} & \begin{tabular}[c]{@{}c@{}}TDNN 128-5-2\\ TDNN 128-3-4\\ Bi-LSTM 128\\ FC 512\end{tabular} \\ \hline
\begin{tabular}[c]{@{}c@{}}Statistics\\ Pooling\end{tabular} & \begin{tabular}[c]{@{}c@{}}256-dim mean \\+ 256-dim std\end{tabular} & \begin{tabular}[c]{@{}c@{}}512-dim mean \\ +512-dim std\end{tabular} \\ \hline
\begin{tabular}[c]{@{}c@{}}Emotion classification\\ sub-network\end{tabular} & FC 512-64-64-4 & FC 1024-128-128-4 \\ \hline
\begin{tabular}[c]{@{}c@{}}Speaker classification\\ sub-network\end{tabular} & FC 512-64-64-8 & FC 1024-128-128-200 \\ \hline
\end{tabular}
}
\label{tab:tab1}
\end{table}

\subsection{Model configuration}

In Table \ref{tab:tab1}, we show the model configurations for the two different data sets in terms of the three sub-networks. Our model is implemented with PyTorch \cite{paszke2017automatic}. For embedding sub-network,  TDNN layer is realized by 1-dimensional convolution, and ``TDNN N-K-D'' (for example ``TDNN 128-5-2'') means 128 output nodes, 5 $\times$ 1 kernel size and dialation is 2. ``Bi-LSTM N'' means we use bidirectional RNN with N LSTM nodes. We also add a FC layer after the output of bidirectional LSTM to increase the feature dimension and avoid too much information loss by statistics pooling on the whole utterance. The dimension is set to 256 for IEMOCAP and 512 for Mandarin speech emotion dataset. We calculat the mean and standard deviation for the statistics pooling step. For emotion and speaker classification sub-network, we use a FC network configured as ``D1-D2-D3-D4'', where D1 means number of input nodes, D2 and D3 for number of hidden nodes in each layer and D4 for number of output nodes. Both TDNN and FC layers (except for output layers) are followed by ReLU activation function, batch normalization (applied to RNN) and dropout with 0.5 keep probability (applied to RNN). We use stochastic gradient descent for optimization, the learning rate of which is set to 1E-03 with Nestrov momentum (factor 0.9). All experiments run 100 epochs, and the epoch with the best performance on validation set is saved for evaluation on testing set. We have four models in total: SER only model, SER and SR in multi-task learning (MTL) setting, DAT setting and CGT setting.

\section{RESULTS AND DISCUSSION}
\label{sec:page}

\begin{table}[]
\caption{Performance comparison on IEMOCAP}
\centering
\resizebox{0.85\columnwidth}{!}{%
\centering
\begin{tabular}{|c|c|c|}
\hline
 & Val & Test \\ \hline
SER\_only & \textbf{57.0\%} ($\pm$ 1.0\%) & 53.9\% ($\pm$ 1.0\%) \\ \hline
SER\_SR\_MTL & 56.6\% ($\pm$ 1.1\%) & 54.9\% ($\pm$ 0.6\%) \\ \hline
SER\_SR\_DAT & 56.6\% ($\pm$ 1.2\%) & 56.5\% ($\pm$ 1.4\%) \\ \hline
SER\_SR\_CGT & 55.6\% ($\pm$ 0.8\%) & \textbf{57.3\%} ($\pm$ 1.2\%) \\ \hline
\end{tabular}
}
\label{tab:tab2}
\end{table}

In Table \ref{tab:tab2}, we show the 4 models' accuracies of classifying 4 emotions in IEMOCAP on both validation and testing set of all 5 folds together. All numbers are the average of 5 times running with standard deviation in parentheses. It can be found that if the model is trained with SER only, it can achieve higher performance on validation set, possibly due to overtuning. However, this gives low accuracy on test set, which means its generalization ability is bad. For MTL, the performance gap between validation set and testing set is smaller, and there is some improvement over single task on testing set. DAT almost produces no performance gap between validation and testing set. CGT achieves the highest accuracy on testing set even with models do not have very high performance on validation set. This verifies that even without adaptation data, DAT can also learn speaker-invariant representation which can generalize to unseen testing speakers. We also observe that CGT can achieve domain generalization in SER on unseen speakers.

\begin{table}[t]
\caption{Performance comparison on Mandarin speech emotion data set}
\centering
\resizebox{0.85\columnwidth}{!}{%
\begin{tabular}{|c|c|c|}

\hline
 & Val & Test \\ \hline
SER\_only & 83.7\% ($\pm$ 0.7\%) & 81.7\% ($\pm$ 1.2\%) \\ \hline
SER\_SR\_MTL & 82.5\% ($\pm$ 0.7\%) & 80.9\% ($\pm$ 0.3\%) \\ \hline
SER\_SR\_DAT & \textbf{84.9\%} ($\pm$ 1.2\%) & \textbf{83.5\%} ($\pm$ 0.6\%) \\ \hline
SER\_SR\_CGT & 82.0\% ($\pm$ 0.6\%) & 81.1\% ($\pm$ 0.3\%) \\ \hline
\end{tabular}
}
\label{tab:tab3}
\end{table}

\begin{figure}[t]
   \centering
   \includegraphics[width=0.6\linewidth]{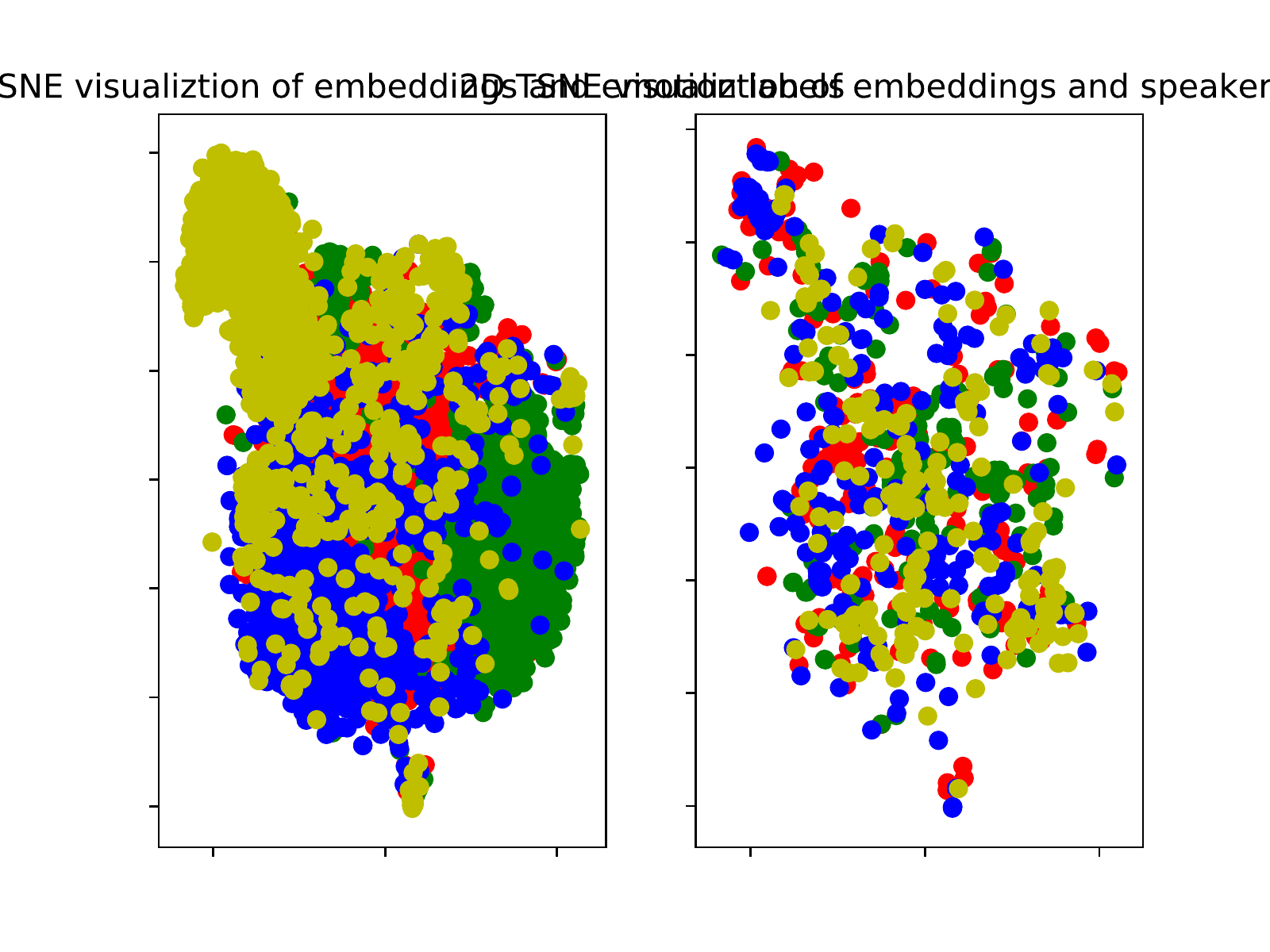} 
   \caption{t-SNE plots of validation set embeddings with both emotion labels (left) and speaker labels (right, we randomly picked 4 speakers).}
   \label{fig:fig2}
\end{figure}

Table \ref{tab:tab3} gives the accuracy comparison among the same 4 models on Mandarin speech emotion data set. All numbers are the average of 3 times running. On the large Mandarin speech emotion data set, a similar trend with IEMOCAP can be observed that the accuracy gap between validation set and testing set is decreasing although for all models, the accuracy on test set is a little worse than on validation set. The MTL model has no gain over single task model. The DAT gives the highest performance on both validation set and test set. It makes sense because when there are a large number of domain labels (speakers in this study), there may exist some training speakers similar with speakers in validation and testing set (target domain). This is equivalent to that DAT is provided with target domain data, thus providing overall improvement of the model. However, CGT is unable to beat the baseline system on this data set. The authors in \cite{shankar2018generalizing} observed similar evaluation results when the size of domain labels is large, and they commented that in this case training data can cover more domain variations, thus the augmentation during training strategy shows no improvement. This explanation aligns with our experimental results: on the small IEMOCAP data set with only 8 speaker labels during training, CGT can yield better generalization ability; On the large Mandarin speech emotion data set with 200 speaker labels during training, it fails to bring benefit. We also show that the embedding $e_i$ generated by our proposed adversarial training is speaker-irrelevant in figure \ref{fig:fig2}. For space consideration, we only show the t-SNE plots \cite{maaten2008visualizing} of learnt $e_i$ by the DAT model. It is obvious that DAT can remove speaker information from embeddings, which allows generalization to new testing speakers.

\section{CONCLUSION}
\label{sec:illust}

In this paper, we propose to use two adversarial training schemes to achieve speaker-invariant representations for speech emotion recognition. The two training strategies are DAT and CGT respectively. While DAT aims to reduce the domain information in representation learning of speech emotion, CGT tackles the problem in a domain generalization way based on domain-guided data augmentation during training. Experiments on a small data set IEMOCAP and a larger Mandarin speech emotion database shows that even without data from target speakers, DAT can still provide gains when testing on new speakers. Although CGT shows on improvement on the larger data set, it still can generalize better than DAT when the size of domain labels is small.


\begin{thebibliography}{10}

\bibitem{chen20183}
Mingyi Chen, Xuanji He, Jing Yang, and Han Zhang,
\newblock ``3-d convolutional recurrent neural networks with attention model
  for speech emotion recognition,''
\newblock {\em IEEE Signal Processing Letters}, vol. 25, no. 10, pp.
  1440--1444, 2018.

\bibitem{luo2018investigation}
Danqing Luo, Yuexian Zou, and Dongyan Huang,
\newblock ``Investigation on joint representation learning for robust feature
  extraction in speech emotion recognition,''
\newblock {\em Proc. Interspeech 2018}, pp. 152--156, 2018.

\bibitem{schuller2010cross}
Bjorn Schuller, Bogdan Vlasenko, Florian Eyben, Martin Wollmer, Andre
  Stuhlsatz, Andreas Wendemuth, and Gerhard Rigoll,
\newblock ``Cross-corpus acoustic emotion recognition: Variances and
  strategies,''
\newblock {\em IEEE Transactions on Affective Computing}, vol. 1, no. 2, pp.
  119--131, 2010.

\bibitem{abdelwahab2017ensemble}
Mohammed Abdelwahab and Carlos Busso,
\newblock ``Ensemble feature selection for domain adaptation in speech emotion
  recognition,''
\newblock in {\em Acoustics, Speech and Signal Processing (ICASSP), 2017 IEEE
  International Conference on}. IEEE, 2017, pp. 5000--5004.

\bibitem{pascual2017segan}
Santiago Pascual, Antonio Bonafonte, and Joan Serra,
\newblock ``Segan: Speech enhancement generative adversarial network,''
\newblock {\em arXiv preprint arXiv:1703.09452}, 2017.

\bibitem{kaneko2017generative}
Takuhiro Kaneko, Hirokazu Kameoka, Nobukatsu Hojo, Yusuke Ijima, Kaoru
  Hiramatsu, and Kunio Kashino,
\newblock ``Generative adversarial network-based postfilter for statistical
  parametric speech synthesis,''
\newblock in {\em Proc. ICASSP}, 2017, vol. 2017, pp. 4910--4914.

\bibitem{hsu2017voice}
Chin-Cheng Hsu, Hsin-Te Hwang, Yi-Chiao Wu, Yu~Tsao, and Hsin-Min Wang,
\newblock ``Voice conversion from unaligned corpora using variational
  autoencoding wasserstein generative adversarial networks,''
\newblock {\em arXiv preprint arXiv:1704.00849}, 2017.

\bibitem{jiao2018simulating}
Yishan Jiao, Ming Tu, Visar Berisha, and Julie Liss,
\newblock ``Simulating dysarthric speech for training data augmentation in
  clinical speech applications,''
\newblock {\em arXiv preprint arXiv:1804.10325}, 2018.

\bibitem{shinohara2016adversarial}
Yusuke Shinohara,
\newblock ``Adversarial multi-task learning of deep neural networks for robust
  speech recognition.,''
\newblock in {\em INTERSPEECH}, 2016, pp. 2369--2372.

\bibitem{sun2017unsupervised}
Sining Sun, Binbin Zhang, Lei Xie, and Yanning Zhang,
\newblock ``An unsupervised deep domain adaptation approach for robust speech
  recognition,''
\newblock {\em Neurocomputing}, vol. 257, pp. 79--87, 2017.

\bibitem{meng2018speaker}
Zhong Meng, Jinyu Li, Zhuo Chen, Yong Zhao, Vadim Mazalov, Yifan Gong, et~al.,
\newblock ``Speaker-invariant training via adversarial learning,''
\newblock {\em arXiv preprint arXiv:1804.00732}, 2018.

\bibitem{wang2018unsupervised}
Qing Wang, Wei Rao, Sining Sun, Leib Xie, Eng~Siong Chng, and Haizhou Li,
\newblock ``Unsupervised domain adaptation via domain adversarial training for
  speaker recognition,''
\newblock in {\em 2018 IEEE International Conference on Acoustics, Speech and
  Signal Processing (ICASSP)}. IEEE, 2018, pp. 4889--4893.

\bibitem{abdelwahab2018domain}
Mohammed Abdelwahab and Carlos Busso,
\newblock ``Domain adversarial for acoustic emotion recognition,''
\newblock {\em arXiv preprint arXiv:1804.07690}, 2018.

\bibitem{ganin2016domain}
Yaroslav Ganin, Evgeniya Ustinova, Hana Ajakan, Pascal Germain, Hugo
  Larochelle, Fran{\c{c}}ois Laviolette, Mario Marchand, and Victor Lempitsky,
\newblock ``Domain-adversarial training of neural networks,''
\newblock {\em The Journal of Machine Learning Research}, vol. 17, no. 1, pp.
  2096--2030, 2016.

\bibitem{snyder2017deep}
David Snyder, Daniel Garcia-Romero, Daniel Povey, and Sanjeev Khudanpur,
\newblock ``Deep neural network embeddings for text-independent speaker
  verification,''
\newblock in {\em Proc. Interspeech}, 2017, pp. 999--1003.

\bibitem{povey2011kaldi}
Daniel Povey, Arnab Ghoshal, Gilles Boulianne, Lukas Burget, Ondrej Glembek,
  Nagendra Goel, Mirko Hannemann, Petr Motlicek, Yanmin Qian, Petr Schwarz,
  et~al.,
\newblock ``The kaldi speech recognition toolkit,''
\newblock in {\em IEEE 2011 workshop on automatic speech recognition and
  understanding}. IEEE Signal Processing Society, 2011, number
  EPFL-CONF-192584.

\bibitem{shankar2018generalizing}
Shiv Shankar, Vihari Piratla, Soumen Chakrabarti, Siddhartha Chaudhuri, Preethi
  Jyothi, and Sunita Sarawagi,
\newblock ``Generalizing across domains via cross-gradient training,''
\newblock {\em arXiv preprint arXiv:1804.10745}, 2018.

\bibitem{zhang2017cross}
Biqiao Zhang, Emily~Mower Provost, and Georg Essl,
\newblock ``Cross-corpus acoustic emotion recognition with multi-task learning:
  Seeking common ground while preserving differences,''
\newblock {\em IEEE Transactions on Affective Computing}, , no. 1, pp. 1--1,
  2017.

\bibitem{busso2008iemocap}
Carlos Busso, Murtaza Bulut, Chi-Chun Lee, Abe Kazemzadeh, Emily Mower, Samuel
  Kim, Jeannette~N Chang, Sungbok Lee, and Shrikanth~S Narayanan,
\newblock ``Iemocap: Interactive emotional dyadic motion capture database,''
\newblock {\em Language resources and evaluation}, vol. 42, no. 4, pp. 335,
  2008.

\bibitem{paszke2017automatic}
Adam Paszke, Sam Gross, Soumith Chintala, Gregory Chanan, Edward Yang, Zachary
  DeVito, Zeming Lin, Alban Desmaison, Luca Antiga, and Adam Lerer,
\newblock ``Automatic differentiation in pytorch,''
\newblock 2017.

\bibitem{maaten2008visualizing}
Laurens van~der Maaten and Geoffrey Hinton,
\newblock ``Visualizing data using t-sne,''
\newblock {\em Journal of machine learning research}, vol. 9, no. Nov, pp.
  2579--2605, 2008.

\end{thebibliography}

\end{document}